\documentclass[twocolumn, showpacs, preprintnumbers, amsmath, amssymb, superscriptaddress, pra]{revtex4}
\usepackage{graphicx}
\usepackage{dcolumn}% Align table columns on decimal point
\usepackage{bm}% bold math
\usepackage{epsf}
\usepackage{amssymb}
\DeclareGraphicsExtensions{.eps}
\usepackage{epstopdf}
\usepackage{natbib}
\usepackage{subfigure}
\usepackage{graphicx}

\begin{document}

\author{David S. Simon}
%\email[e-mail: ]{simond@bu.edu}
\affiliation{Dept. of Physics and Astronomy, Stonehill College, 320 Washington Street, Easton, MA 02357}

\affiliation{Department of Electrical and Computer Engineering, Boston
University, 8 Saint Mary's St., Boston, MA 02215, USA}

\author{Gregg Jaeger}

\affiliation{Department of Electrical and Computer Engineering, Boston
University, 8 Saint Mary's St., Boston, MA 02215, USA}

\affiliation{Division of Natural Sciences and Mathematics, Boston University,  Boston,
MA 02215, USA}
%\email[e-mail: ]{jaeger@math.bu.edu}

\author{Alexander V. Sergienko}
%\email[e-mail: ]{alexserg@bu.edu}
\affiliation{Department of Electrical and Computer Engineering, Boston
University, 8 Saint Mary's St., Boston, MA 02215, USA}

\affiliation{Photonics Center, Boston
University, 8 Saint Mary's St., Boston, MA 02215, USA}

\affiliation{Dept. of Physics, Boston University, 590 Commonwealth
Ave., Boston, MA 02215, USA}

\begin{abstract} An entanglement witness approach to quantum coherent state key distribution and a system
for its practical implementation are described. In this approach, eavesdropping can be detected by
a change in sign of either of two witness functions, an entanglement witness ${\cal S}$ or an
eavesdropping witness ${\cal W}$. The effects of loss and eavesdropping on system operation are
evaluated as a function of distance. Although the eavesdropping witness ${\cal W}$ does not
directly witness entanglement for the system, its behavior remains related to that of the true
entanglement witness ${\cal S}$. Furthermore, ${\cal W}$ is easier to implement experimentally than
${\cal S}$. ${\cal W}$ crosses the axis at a finite distance, in a manner reminiscent of
entanglement sudden death. The distance at which this occurs changes measurably when an
eavesdropper is present. The distance dependance of the two witnesses due to amplitude reduction
and due to increased variance resulting from both ordinary propagation losses and possible
eavesdropping activity is provided. Finally,  the information content and secure key rate of a
continuous variable protocol using this witness approach are given.

\end{abstract}


\title{Coherent State Quantum Key Distribution with Entanglement Witnessing}

\maketitle

\section{Introduction}\label{introsection}

The goal of quantum key distribution (QKD) is for two participants (Alice and Bob) to generate a
shared cryptographic key of bits in such a way that quantum mechanics prevents an eavesdropper
(Eve) from obtaining significant information about the key without being detected. QKD schemes \cite{bb84,e91}
based on the transmission of single photons or entangled photon pairs tend to be highly secure \cite{shor}.
However, because single photons can be easily absorbed or deflected, the operational distances and
key generation rates of these schemes are limited. It is often desirable to instead use pairs of entangled
coherent states, because individual-photon-level losses have little effect on them. Along with this
benefit comes the challenge of revealing the action of eavesdroppers: it suffices for Eve to obtain only a
small fraction of the coherent state beam to measure the transmitted state. Moreover, although
pairs of entangled coherent states can be created \cite{sanders1,sanders2}, randomly modulating
them as needed for QKD is a nontrivial task.

Recently \cite{kirby}, a technique applicable to detection of an eavesdropper on a quantum optical
communication channel was proposed which involved phase-entangling two coherent state beams by
interaction with a single photon inside a nonlinear medium. In that scheme, a beam splitter first
puts a photon into a superposition of two possible path states. A phase shift is induced conditionally,
depending on the path state, so that the pair of beams becomes phase-entangled. Alice and Bob each
receive one beam and make homodyne measurements to determine its phase. The relative phase between
the beams determines the bit value to be used in the key. Effects due to eavesdropping are made
detectable by introducing additional interferometers with controllable phase shifts $\sigma_1$ and
$\sigma_2$ just before each of the detectors, respectively. Interference terms then appear in the
joint detection rate as $\sigma_1$ and $\sigma_2$ are varied. If the beams have not been disturbed
in transit, the visibility of this interference should be greater than ${1\over \sqrt{2}}\approx
70.7\% $, suggesting stronger-than-classical correlations and violation of a Bell-type inequality.
If the visibility drops below $70.7\%$, this could indicate that the beam has been tampered with.
This method in principle allows phase-entangled states to be robustly distributed over large
distances.

In this paper, we propose a new technique for revealing eavesdroppers in systems for quantum key
bit distribution. This technique introduces entanglement in a manner similar to \cite{kirby}, but
uses a fundamentally different approach to eavesdropper detection. Rather than using Bell violation
for checking security, the idea is to instead look for degradation or death of entanglement due to
Eve's actions by using functions designed to witness it \cite{horod1,duan}. The switch from
measurements of non-local interference associated with a Bell-type inequality direct
entanglement-related witnesses provides substantial benefits: it both expands the effective
operating distance and simplifies the required apparatus. The increase in operating distance is due
to the fact that Bell violation is a stronger condition than entanglement. The particular entanglement witness
${\cal S}$ \cite{shchukin} used is negative for all finite distances when the coherent states
propagate undisturbed; however, ${\cal S}$ changes sign to a positive value in the presence of
eavesdropping, thus revealing Eve's intervention. Another related witness function ${\cal W}$ which
is more easily measured but does not directly indicate entanglement in our system can also serve
this purpose. To our knowledge, this is the first time such an approach has been proposed for use
in QKD.

As in \cite{kirby}, which involves the Bell inequality, the main goal of these functions is simply
to reveal the presence of eavesdropping on the line; when the eavesdropper's signature is observed,
the communicating parties know to shut down the line and seek another communication channel. The
actual bits either may be derived from the entangled phases or they may arise from normal telecom
approaches of modulating the intense coherent states. In this sense, the goal is to provide a
``quantum tripwire'' for practical use, as opposed to absolute security in the sense that the
phrase is commonly used in QKD. In other words, the basic idea is to take a more pragmatic approach
to communication by providing an extra quantum-based layer of security to support highly efficient
classical communication. As a result, our primary goal is less general and less difficult to
achieve than other continuous variable protocols
\cite{ralph,hillary,reid,cerflevy,grosshans,weedbrook,garcia} that have been proposed with the goal
of unconditional security in mind.  Nonetheless, as discussed in Sec. \ref{keysection} the witness approach
is used directly on the key-bit transmitting system to provide security to fully quantum
communication as well.

It is because the witness ${\cal S}$ itself involves third-order correlation functions, which may
be inconvenient to implement experimentally, that we also consider the second witness function
${\cal W}$. ${\cal W}$ is related by rescaling of the quadratures to a well-known entanglement
witness ${\cal W}_s$ \cite{rsimon,barbosa}, but is not in the strict sense a true entanglement
witness in the current context. Despite this, it gives eavesdropper-detection results that match
well with those of ${\cal S}$, and has the additional advantage that it is built from the
covariance matrix of the system, which is easily accessible experimentally. ${\cal W}$ starts from
an initially negative value, but then crosses the axis to positive values at finite distance, both
during free propagation and in the presence of eavesdropping. This is closely analogous to the
phenomenon of entanglement sudden death (ESD) \cite{ann}, in which entanglement is lost after
propagating a finite distance. The crossing occurs at a distance that can be easily predicted when
there is no eavesdropping present. When eavesdropping occurs, the curve of ${\cal W}$ versus
distance shifts by a measurable amount; in particular, there is a clear alteration of the distance
at which the sign changes, allowing for easy detection.

We will collectively refer to quantities which are measurably altered by predictable amounts in the
presence of eavesdropping as eavesdropping witnesses; both the true entanglement witness ${\cal S}$
and the additional function ${\cal W}$ are examples of such functions. It is shown that the two
give consistent results for the distance over which the entanglement becomes unusable for
eavesdropper detection.

Throughout this paper, coherent state quadratures will be defined in terms of creation and
annihilation operators via the relations \begin{equation}\hat q = {1\over 2}\left( \hat a +\hat
a^\dagger \right),\qquad \hat p = {1\over {2i}}\left( \hat a -\hat a^\dagger \right).\end{equation}
It should be noted that there are several other normalization conventions that are common in the
literature, with different constants in front on the right-hand side. Accordingly, when results
from other authors are quoted in the following sections, the form used here may differ from their
originally published forms by factors of two in some terms.

We begin in Section \ref{phasesection} by describing the entangled states under consideration and
their means of production. The eavesdropping model assumed is described in Sec. \ref{clonesection}.
There, we model the eavesdropping procedure by introducing a Gaussian cloner into the path of one of the
coherent states. We then introduce the entanglement witness ${\cal S}$ and analyze its behavior in
Section \ref{entwitsection}. In order to have a more convenient experimental measure, we then
introduce ${\cal W}$ in \ref{evewitsection}, and look in section
\ref{thresholdsection} at some of its properties, with emphasis on its behavior under eavesdropping. Discussion of some
information-related aspects in Sec. \ref{keysection} is then followed by a brief discussion of the
results in section \ref{conclusion}.

\section{Phase-Entangled Coherent States}\label{phasesection}

\begin{figure}%[htb]
\begin{center}
\subfigure[]{
  \includegraphics[width=.7\columnwidth]{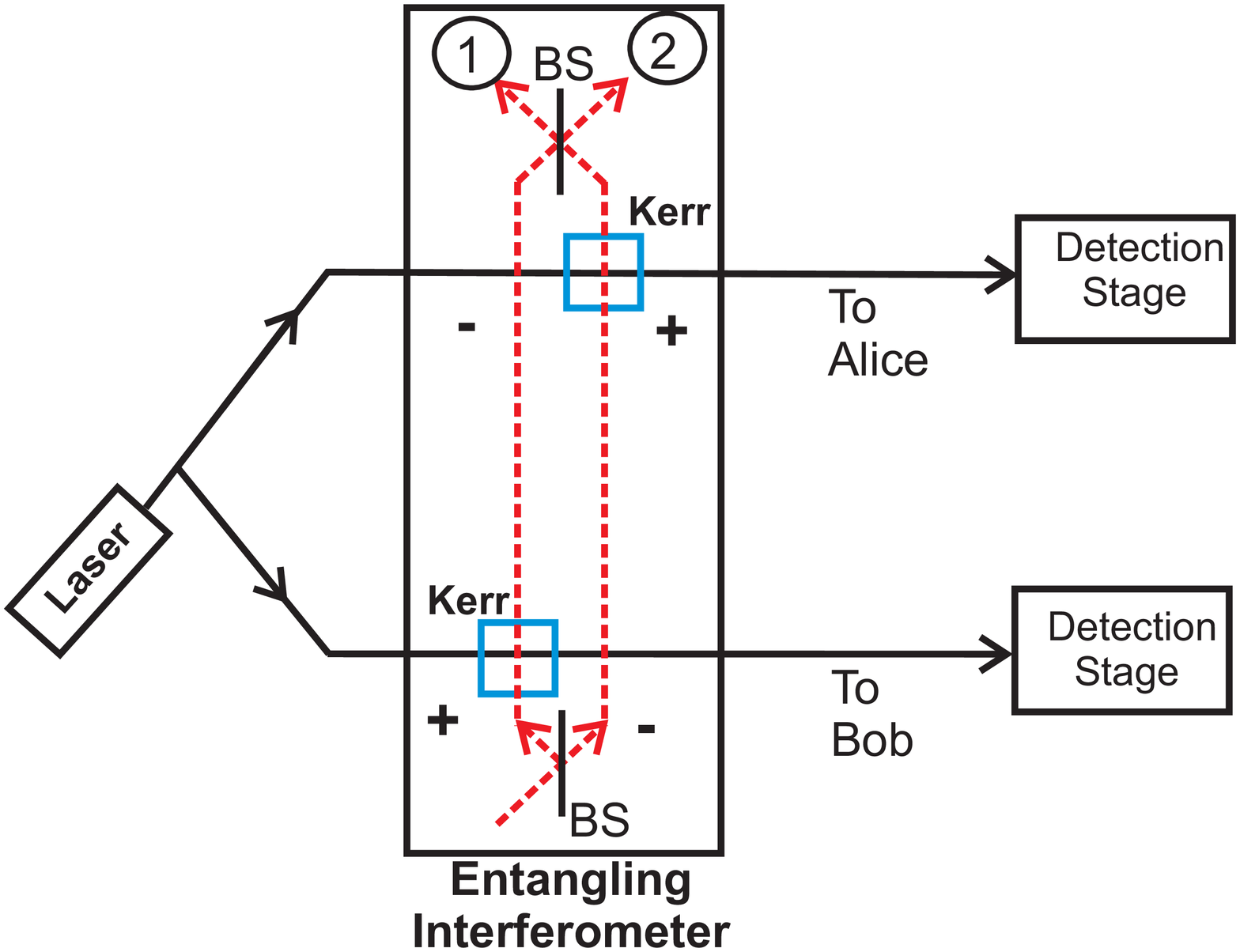}}
\subfigure[]{
\includegraphics[width=.7\columnwidth]{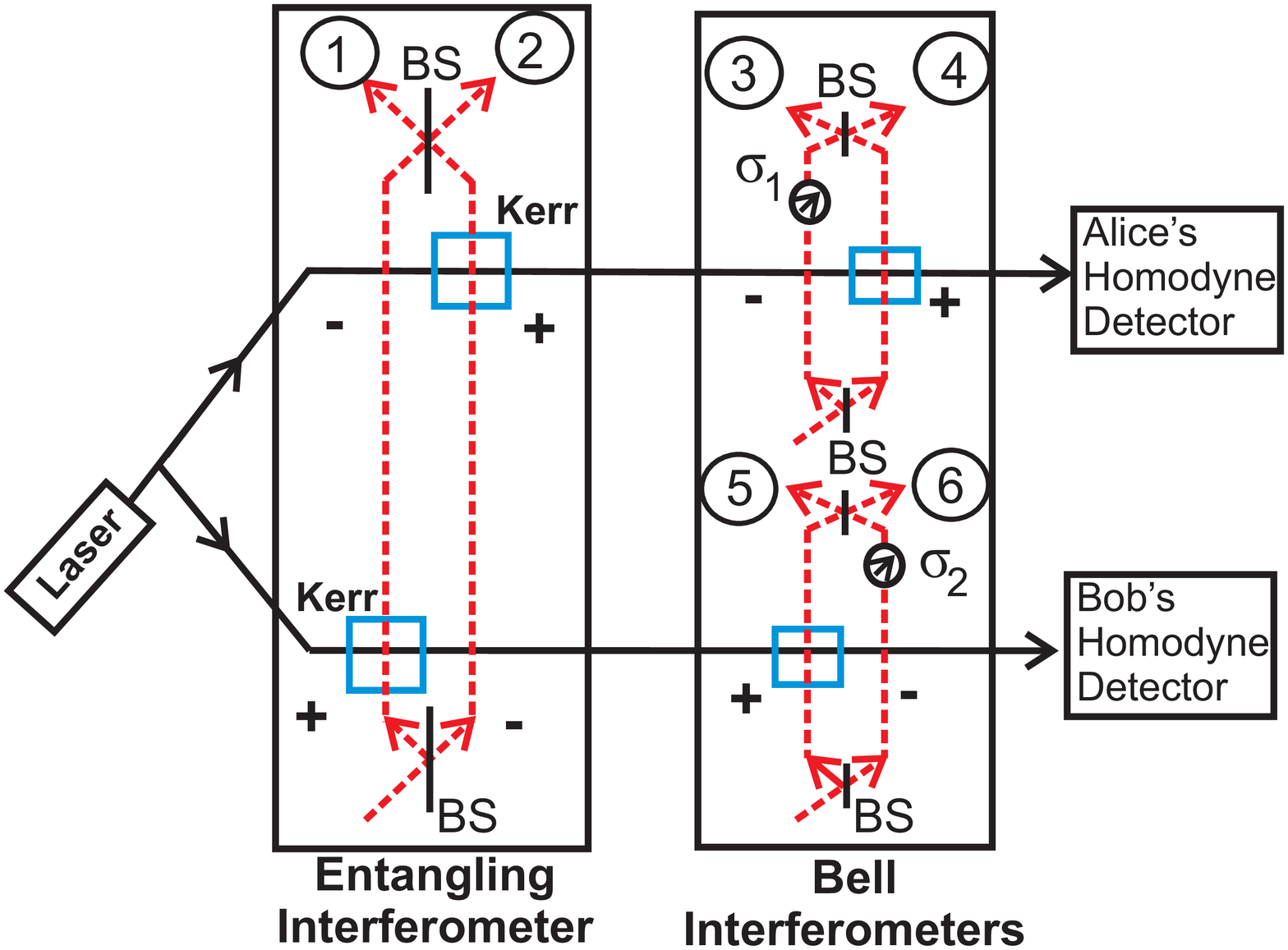}}
\caption{(Color online) Schemes for phase-based coherent state key distribution with
single-photon triggers. (a) Scheme of the current paper. A beam splitter splits a laser beam into two beams in identical coherent
states (solid black lines); a phase shifter compensates for the phase gained in the reflected
state. A single photon also enters a superposition of two path states (dashed red lines). Due to
the joint interaction of coherent state and the photon within Kerr media, the beams enter an
equal-weight superposition of product states of pairs of oppositely phase-shifted coherent states.
The specific form of the detection unit will be different for each of the applications to be discussed in the text.
(b) Scheme of \cite{kirby}, with two additional interferometers to test for Bell violations.}
\label{setupfig}
\end{center}
\end{figure}

The apparatus for the proposed system is shown in Fig. \ref{setupfig} (a). A laser followed by a
beam splitter produces a pair of optical coherent states, each in state $|\alpha\rangle$.  As in
\cite{kirby}, the coherent state subsystem pair initially produced in state $|\alpha\rangle_A
|\alpha \rangle_B $ becomes entangled in an interferometer by coupling to a single photon. A beam
splitter first causes the photon state to enter a superposition of two path eigenstates. Then if
the photon is in the upper path state, beam $B$ gains a phase shift $2\phi$ due to cross-phase
modulation of the photon with that beam in a nonlinear Kerr medium
\cite{nemoto,munro,lukin,harris,schmidt,Turchette}, whereas if the photon is in the lower path
state, then there is a phase shift of $2\phi$ in beam $A$. Finally, by adding another constant
phase shift to each beam, we can then arrange the output to be in the entangled state
\begin{equation}|\psi\rangle = {N\over \sqrt{2}}\left( |\alpha_+\rangle_A |\alpha_-\rangle_B +
e^{i\theta} |\alpha_-\rangle_A |\alpha_+\rangle_B \right) ,\label{psistate}\end{equation} where
\begin{equation}|N|^{-2}=\left( 1+\cos\theta \; e^{-4|\alpha|^2\sin^2\phi}\right)\label{initialnorm}\end{equation} and
$\alpha_\pm \equiv \alpha e^{\pm i\phi}$. (For simplicity, we do not explicitly indicate the
single-photon states.) In the following, operators with subscripts $1$ and $2$, respectively, will
correspond to Alice's beam and to Bob's.

Note that, whereas $\pm\phi$ are the phase shifts of the coherent states within a given path state,
$\theta$ is the relative phase between the two joint path states of the photon. The value of joint
phase $\theta$ can be controlled by the experimenters: Keeping only events in which the photon is
detected at detector 1 leads to $\theta =\pi$, while events in which it exits at detector 2 lead to
$\theta =0$. (Other values of $\theta$ can be achieved if desired by, for example, putting a piece
of glass in one of the potential single photon paths.) If the interferometer lacks stability,
randomly-varying phases in the single-photon paths could lead to decoherence. But these photons
could be kept on a single bench in Alice's lab and be well-controlled to prevent this. Fluctuations
in the phases  of the coherent states $|\alpha_\pm\rangle_A |\alpha_\mp\rangle_B \to |\alpha_\pm
e^{ i\delta\phi_1(t)}\rangle_A |\alpha_\mp e^{ i\delta\phi_2(t)}\rangle_B$ would be a more serious
problem because these are shared between labs that may be widely separated. This random phase
variation is an independent source of entanglement loss, separate from the entanglement loss due to
amplitude decay and eavesdropping. (We focus here on the latter, leaving the former to be discussed
elsewhere.)

Using homodyne detection, each participant can measure the phase of his or her beam to determine the sign of its shift.
Because the shifts in the two beams are always opposite, this is sufficient for Alice and Bob to obtain common key bits;
for example, if Alice has $+\phi$ and Bob has $-\phi$, they can take the common bit value to be $0$, while the opposite
case then corresponds to $1$.

Unfortunately, an eavesdropper may extract part of the beam and determine the bit transmitted.
Although this cannot be \emph{prevented}, it can be \emph{detected}, so that Alice and Bob can
prevent key material from being compromised by shutting down the communication line. Recall that,
for the purpose of revealing Eve's intervention, the proposal of \cite{kirby} is to include two
additional interferometers (Fig. \ref{setupfig} (b)), each coupling one beam to another photon in
order to detect nonlocal interference for Bell inequality tests. That approach has at least two
limitations: (i) On the theoretical side, detecting Eve only requires \emph{entanglement}, which in
practice may still exist even when the Bell inequality is not violated \cite{Werner}; thus, the
setup tests for a less than ideal property. (ii) On the experimental side, simultaneous
single-photon events are needed in {\em three independent interferometers}. This low-probability
triple-coincidence in widely-separated interferometers is a significant practical limitation. The
method given in the present paper avoids this problem by removing the need for more than one
interferometer.

Because the amplitude of the input beam can be easily tuned, the system can be adjusted to work at
different operating distances, potentially (as we see in the following sections) up to distances of
several hundred kilometers. Current technology can realistically reach amplitudes $|\alpha |$ of up
to $10^3 -10^4$ without doing damage to the fibers or producing high amounts of fluorescence and
scattering; but for illustrative purposes of future potential we have included plots with values of
up to $10^6$ at some points in the following.

\section{The Effect of Eavesdropping}\label{clonesection}

\begin{figure*}%[ht]
\subfigure[]{
\includegraphics[width=.9\columnwidth] {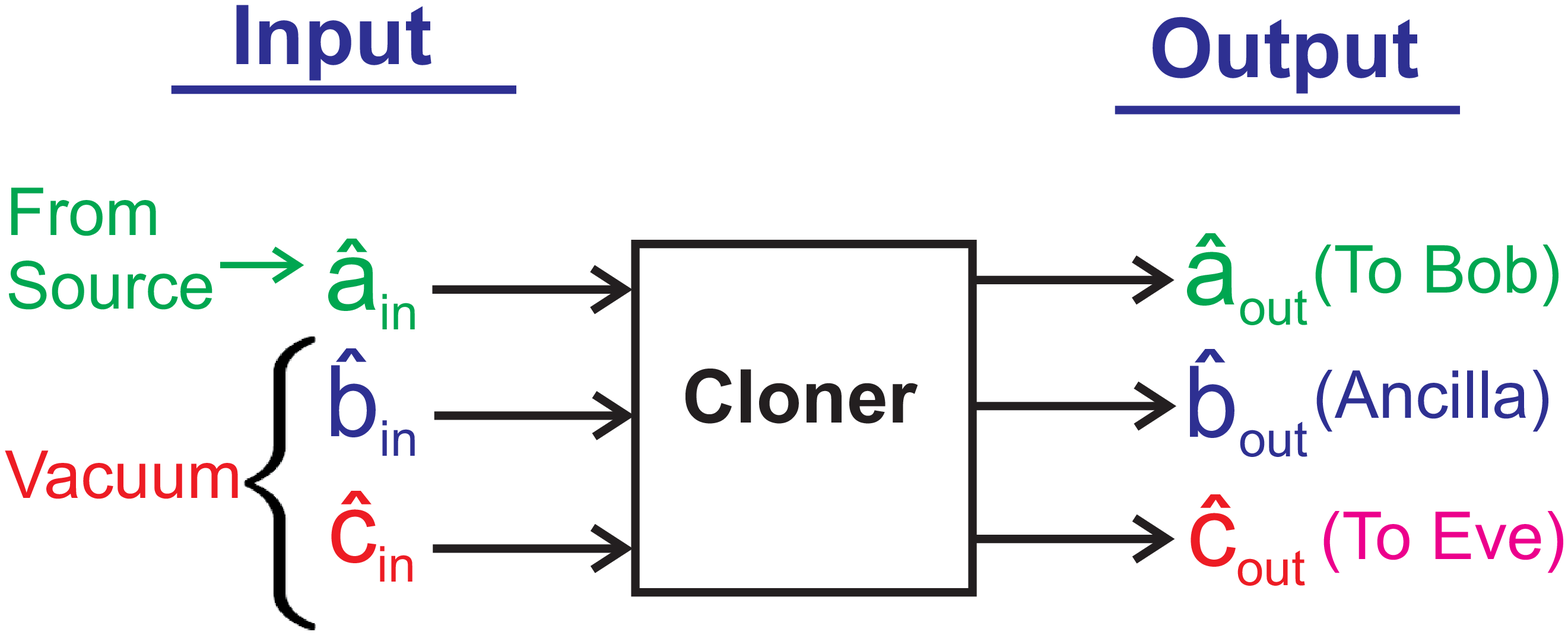}}
\qquad \subfigure[]{
\includegraphics[width=.9\columnwidth]{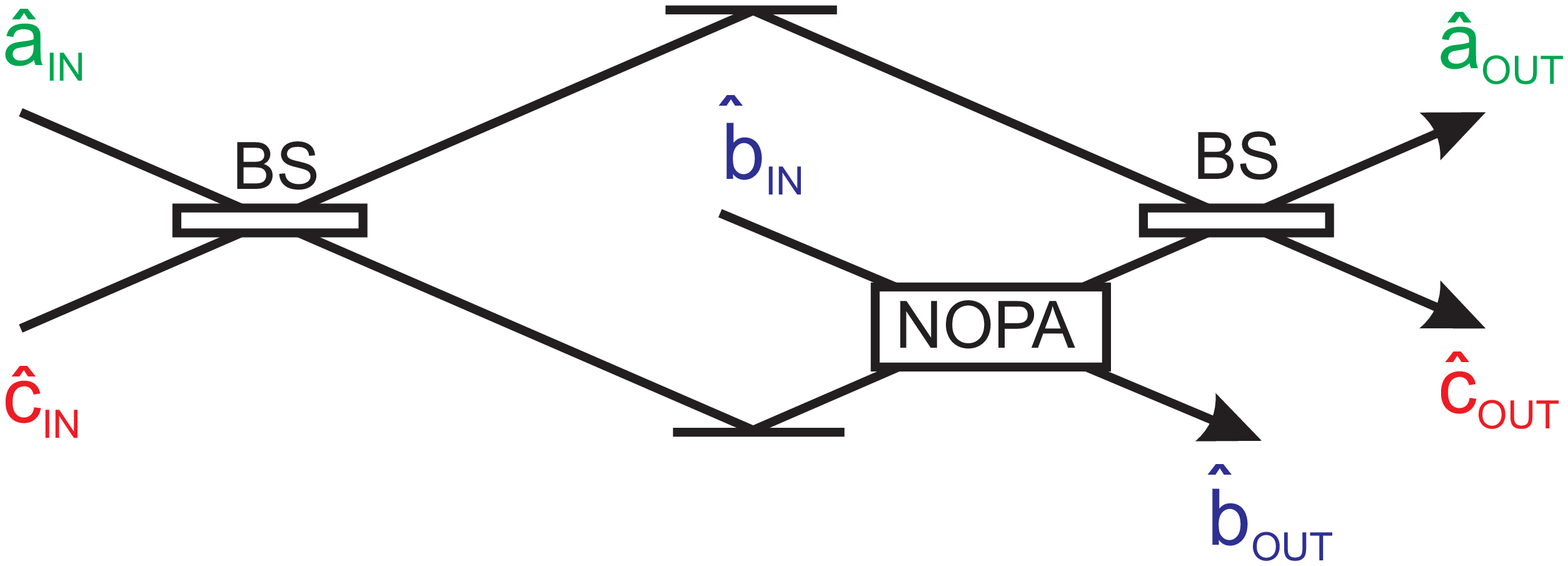}}
\caption{(Color online) A model of a Gaussian cloner \cite{fiurasek} applied by Eve to Bob's beam. The cloner can be realized
by combining an amplifier with a beam splitter. Besides the input from the source ($a_{in}$), there are two
additional inputs: one to the amplifier ($b_{in}$), the other to the first beam splitter ($c_{in}$). The result is two
outputs with quadratures that have means equal to that of the input. One copy ($a_{out}$) is sent
on to Bob. Eve keeps the other ($c_{out}$) to make measurements on.}
\label{clonefig}
\end{figure*}

To examine measures against eavesdropping, we consider the case in which Eve
attaches a  Gaussian cloner \cite{cerf} to one of the beams, which we assume to be Bob's. The
cloner takes an input beam and makes two copies that have the same mean amplitude as the input. Eve
keeps one beam and sends the other on to Bob. But inevitably, there is a net increase in the
variance of Bob's beam that will indicate her presence. Moreover, the more exact a copy Eve's beam
is (i.e. the lower its variance), the larger the disturbance to Bob's beam. Specifically, if
$\sigma_{Bj}$ and $\sigma_{Ej}$ (for $j=q,p$) are the added variances to Bob's beam and to Eve's,
in excess of the initial variance, then these variance increases must satisfy \cite{cerf}:
\begin{equation}\sigma_{Bq}^2\sigma_{Ep}^2\ge {1\over {16}},\qquad
\sigma_{Bp}^2\sigma_{Eq}^2\ge {1\over {16}}.\end{equation} For optimal cloning devices, the effect
on the $q$ and $p$ quadratures should be the same; henceforth, we therefore assume that
$\sigma_q^2=\sigma_p^2\equiv \sigma^2$ for all participants.

In addition to the increased variance, any cloning device will involve additional input ports
besides the one carrying the state to be cloned. These will introduce additional unmeasured
fluctuations, converting a pure input state into a mixed output state \cite{cerf}, consequently
leading to a loss of coherence between previously entangled states. We consider eavesdropping on
only one of the two channels because, given our emphasis on eavesdropper detection, this is the most
advantageous situation for Eve: placing cloners
in both channels can only make make her situation worse by affecting Alice's state as well.

A generic schematic of a Gaussian cloner is shown in Fig. \ref{clonefig}(a). In addition to the
input beam to be cloned (represented by annihilation operator $\hat a_{in}=\hat a_2$), there is an
input $\hat c_{in}$, assumed to be in a vacuum state, onto which the clone is to be imprinted at
output. One further input port $\hat b_{in}$ leads to an internal amplifier. We assume the specific
model of Ref. \cite{fiurasek}, realized in terms of two beam splitters and a nondegenerate optical
parametric amplifier (NOPA), as in Fig. \ref{clonefig}(b). There are three output beams: an ancilla
($\hat b_{out}$) and two clones of the input state. One clone ($\hat a_{out}$) is sent on to Bob,
and one ($\hat c_{out}$) is kept by Eve. The input-output relations for the operators in the
Heisenberg picture are \cite{fiurasek}
\begin{eqnarray} \hat a_{out} &=&   \hat a_{in} -{{e^{-\gamma}}\over\sqrt{2}}\left( \hat c_{in} +\hat b_{in}^\dagger\right)\\
\hat b_{out} &=&   -\sqrt{2}\sinh\gamma \hat c_{in}^\dagger +\sqrt{2}\gamma \hat b_{in} -\hat a_{in}^\dagger \\
\hat c_{out} &=&   \hat a_{in} +{{e^{+\gamma}}\over\sqrt{2}}\left( \hat c_{in} -\hat b_{in}^\dagger\right) .
\end{eqnarray} Here, the asymmetry between the two clones is measured by a parameter $\xi$ which has value
$\xi ={{\ln 2}\over 2}$ for the symmetric case. Then $\gamma =\xi -{{\ln 2}\over 2}$ measures the
deviation from symmetry. The optimal case of $\gamma =0$ produces fidelity $F_a=F_c={2\over 3}$ for
both clones. It is readily verified that the mean values at both outputs are unchanged from the
input, $\langle \hat q_E \rangle=\langle \hat q_2^\prime \rangle =\langle \hat q_2 \rangle$ and
$\langle \hat p_E \rangle=\langle \hat p_2^\prime \rangle =\langle \hat p_2 \rangle$. It is also
straightforward to show that the variances satisfy
\begin{eqnarray} \Delta q_{a,out}^2 &=& \Delta q_{a,out}^2 +{1\over 4}e^{-2\gamma}\label{varq}\\
\Delta p_{a,out}^2&=& \Delta p_{a,out}^2 +{1\over 4}e^{-2\gamma} \label{varp}
\end{eqnarray} for the clone sent to Bob, and \begin{eqnarray} \Delta q_{c,out}^2 &=& \Delta q_{c,out}^2
+{1\over 4}e^{+2\gamma}\label{vareq}\\
\Delta p_{c,out}^2 &=& \Delta p_{c,out}^2 +{1\over 4}e^{+2\gamma}\label{varep}
\end{eqnarray} for the clone kept by Eve.
Due to the cloning procedure, Bob and Eve each therefore gain added variances (beyond the original
variance of the beam in transit to Bob)  of $\sigma_B^2={1\over 4}e^{-2\gamma}$ and
$\sigma_E^2={1\over 4}e^{2\gamma}$, respectively.

In the Schr\"odinger picture, the cloner has the effect of altering the state: a pure input state
will be converted to a mixed output with a probability distribution of width $\sigma_B^2$
\cite{cerf}, which will inevitably damage or destroy the entanglement of the cloned state with
Alice's state.

\section{Entanglement Witness Approach.}\label{entwitsection}

Recall that, using the Bell--CHSH inequality, the absolute value of the expectation value of the
Bell--CHSH operator ${\mathcal B}$, when properly applied, provides a necessary and sufficient
indication of the presence or absence of entanglement for pure states. In that sense, the absolute
value  $|{\mathcal B}|$ is the longest-used strong entanglement witness. Here, in place of
$|{\mathcal B}|$ falling below the critical Bell inequality value 2 as the indicator of loss of
entanglement, we use the loss of the negative-valuedness of an entanglement witness
${\mathcal S}$ that is observable with
a much simpler apparatus. To our knowledge, this is the first time the use of an entanglement
indicator other than the expectation value of a Bell-type operator has been proposed for use in
entangled coherent state QKD.

An entanglement witness is a quantity which is negative whenever a system is entangled; in general,
when it is non-negative this is no longer the case and nothing can be said about the entanglement or separability of the system. Entanglement witnesses can often be based on the positive partial trace (PPT) criterion of
\cite{peres,horod2}. For continuous variables, the most common such witnesses are formed from the
second-order correlation functions (i.e. on covariance matrices). These are extremely useful
because Gaussian states are completely determined by their means and covariance matrices; as a
result, such witnesses often completely characterize the entanglement properties of Gaussian
states. In particular, some entanglement witnesses, such as the function ${\cal W}_s$  mentioned in
section \ref{evewitsection}, are both {\em necessary} and {\em sufficient} conditions for
entanglement when applied to Gaussian states, being positive if {\em and only if} the state is
separable. Such witnesses are referred to as strong witnesses.

However, covariance-based entanglement measures, which do not take into account correlations among
higher moments, may not be fine enough a measure to detect entanglement in non-Gaussian systems, so
a number of higher-order entanglement measures have been discussed in the literature
\cite{mancini,agarwal,raymer,shchukin,gomes}. These involve expectation values of operators formed
from products of more than two creation or annihilation operators (or, equivalently, products of
more than two quadrature operators). Here we will consider one such measure, denoted ${\cal S}$,
and show that it can detect the presence of eavesdropping: when an eavesdropper acts, it will
switch sign from negative to positive values. Because ${\cal S}$ is only a necessary and not a
sufficient measure for entanglement - in other words, it is not a strong witness - it cannot be
said with certainty that entanglement is lost when the sign changes. Whether or not entanglement
persists after the sign change is ultimately beside the point for our current purpose: the sign
change in any case indicates the presence of an eavesdropper, which is our goal. In addition, so
long as the sign does remain negative we \emph{can} say with certainty that the system remains
entangled, and that under an appropriate protocol it therefore remains secure. If ${\cal S}<0$ then
entanglement persists and communication can continue; but if ${\cal S}\ge 0$, communication should
be shut down in order to assure security, even though there is a chance that entanglement still
persists.

The entanglement witness to be used here was introduced in \cite{shchukin} and is defined by the
determinant
\begin{equation}{\cal S}=\left| \begin{array}{ccc}1 & \langle \hat a_2^\dagger\rangle &  \langle \hat a_1 \hat
a_2^\dagger\rangle\\ \langle\hat a_2\rangle & \langle \hat a_2^\dagger \hat a_2\rangle & \langle
\hat a_1 \hat a_2^\dagger \hat a_2\rangle \\ \langle \hat a_1^\dagger \hat a_2\rangle & \langle
\hat a_1^\dagger \hat a_2^\dagger \hat a_2\rangle & \langle \hat a_1^\dagger \hat a_1 \hat
a_2^\dagger \hat a_2\rangle . \label{Sdef}\end{array} \right|\end{equation} Here $\hat a_1$ is the
annihilation operator at Alice's location and $\hat a_2$ is the corresponding operator for Bob's.
This witness is valid for any state, Gaussian or otherwise, and when it is negative the state is
guaranteed to be entangled. Because ${\cal S}$ involves third-order correlations in addition to
second and fourth order, it is more difficult to measure experimentally, although such measurements
have been done \cite{dahlstrom}. The only change of the setup from Fig. \ref{setupfig}(a) is that
the homodyne detectors would be replaced by a more complex detection unit.

Given the explicit form of the entangled bipartite coherent state $|\psi\rangle$ given in Eq.
(\ref{psistate}), ${\cal S}$ can be readily calculated. We find the elements of the matrix at zero
distance are
\begin{eqnarray}\langle \hat a_2^\dagger \rangle &=& \langle \hat a_2 \rangle \;= \; \alpha
\cos\phi \label{moment1}\\
\langle \hat a_1 \hat a_2^\dagger \rangle &=& \langle \hat a_1^\dagger \hat a_2 \rangle \nonumber \\
&=&  \alpha^2|N|^2
\left( \cos 2\phi +e^{-4\alpha^2\sin^2\phi}\right)\label{moment2} \\
\langle \hat a_2^\dagger \hat a_2\rangle &=& \alpha^2|N|^2 \left( 1+\cos 2\phi
e^{-4\alpha^2\sin^2\phi}\right) \label{moment2a} \\ \langle \hat a_1\hat a_2^\dagger \hat a_2\rangle &=& \alpha^3\cos\phi \label{moment3}\\
\langle \hat a_1^\dagger \hat a_1 \hat a_2^\dagger \hat a_2\rangle &=&\alpha^4 .\label{moment4}
\end{eqnarray} It is straightforward to verify that ${\cal S}\to 0$ as $\alpha\to 0$ or $\alpha\to\infty$,
while ${\cal S}<0$ at all finite values of $\alpha$. All terms in the determinant are proportional
to $\alpha^6$, with additional amplitude dependence coming from the exponential terms in Eqs.
(\ref{moment2}) and (\ref{moment3}); the latter terms are negligible except when $\alpha<<1$. For
small $\phi$, the terms in ${\cal S}$ nearly cancel, leaving ${\cal S}$ with a small (negative)
value. ${\cal S}\to 0$ continuously as $\phi\to 0$, i.e. as the state becomes separable.

Distance dependance can be taken into account by replacing the amplitude in each arm by $\alpha\to
\alpha_j(d_j)=\alpha t_j(d_k)$, where $t_j$ is a transmission function in the $j$th branch, for
$j=1,2$. We assume that $\phi<<1$ and $\alpha\phi>>1$ initially, but due to losses $\alpha$ will
eventually decay to small values, at which point the phase space regions centered at $\alpha e^{\pm
i\phi}$ may begin to overlap, resulting in entanglement loss. For propagation losses alone, the
transmission functions are of form $t_j (d_j)=e^{-{1\over 2}K_j d_j}$, with propagation distance
$d_j$ in each arm.  When these losses are included, expressions of the form $\langle \hat a_1^l\hat
a_1^{\dagger m} \hat a_2^n\hat a_2^{\dagger p}\rangle$ are multiplied by factors of $e^{-{K\over
2}\left[(l+m)d_1-(n+p)d_2\right]}$, while the exponential terms in Eqs.
(\ref{moment1})-(\ref{moment4}) and in the normalization constant Eq. (\ref{initialnorm}) become
$\exp\left( {-4\alpha^2e^{-{K\over 2}(d_1+d_2)}\sin^2\phi}\right) .$ Given this, the entanglement
witness can be calculated as a function of distance for various parameter values.

\begin{figure}[ht]
%\subfigure[]{
%  \includegraphics[width=.9\columnwidth]{splotlong}}
%\subfigure[]{
%  \includegraphics[width=.9\columnwidth]{splot1}}
%\subfigure[]{
%  \includegraphics[width=.9\columnwidth]{splot2}}
\begin{center}
\includegraphics[width=.9\columnwidth]{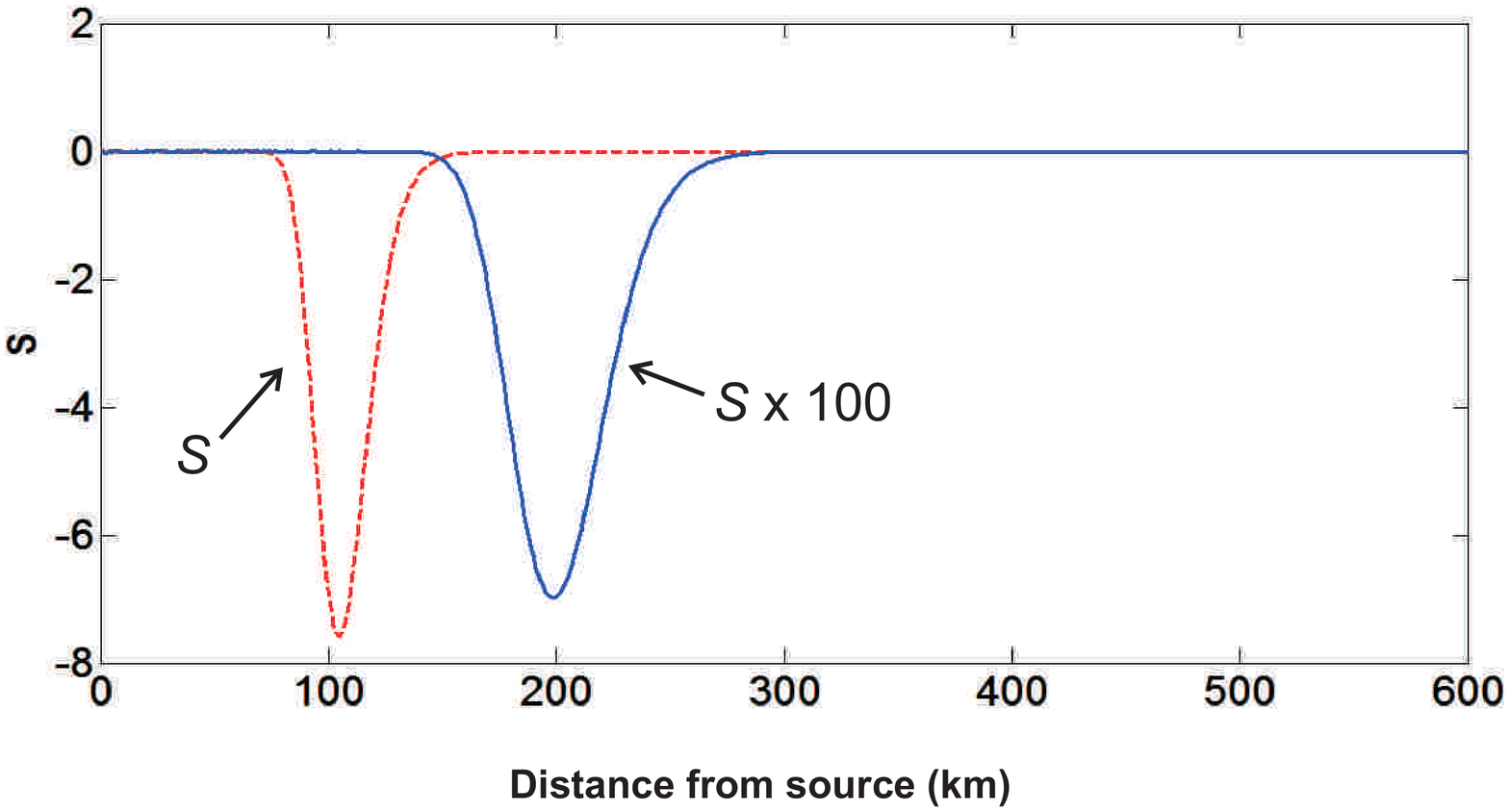}
\caption{(Color online) (a) Behavior of entanglement witness ${\cal S}$ as a function of distance, assuming that
the amplitudes have decay constants $K=.046 \; {\mbox km}^{-1}$. Here, $\alpha =100$ and $\phi =.1$. The red
dashed line assumes symmetric decay. The solid blue line assumes that the source is in Alice's lab, so that decay occurs only on one side; the values in this
latter case were magnified by a factor of 100 before plotting.
%(b) Blown up views of the large negative dip for $\phi=.1$, assuming symmetric decay in the two beams. From left to right, the %initial amplitudes are respectively
%$\alpha =100,\; 500 , \; 1000$. (c) Same as part b, but now assuming that the source is in Alice's lab, so there is decay only %on Bob's side.
}\label{svsdistfig}
\end{center}\end{figure}

\begin{figure}%[htb]
\begin{center}
\includegraphics[scale=.30]{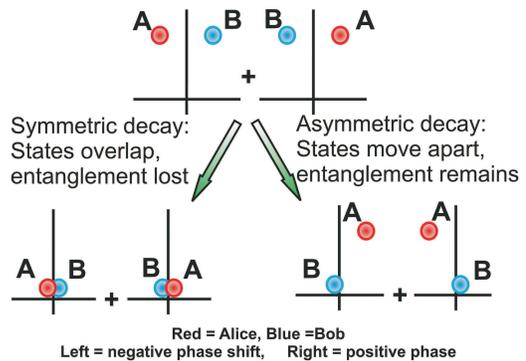}
\caption{(Color online) The larger distance
of disentanglement for the asymmetric case in Fig. \ref{svsdistfig} is due the fact that the two coherent states move apart in phase
space, whereas in the symmetric case both decay toward the same vacuum state.}
\label{ESDfigb}
\end{center}
\end{figure}

Plots of ${\cal S}$ versus distance are shown in Fig. \ref{svsdistfig} for several parameter
values. Two cases are shown: the case of equal decay in both arms (Alice and Bob equal distances
from the source), and for decay in one arm only (Alice acting as the source). Note that for the
asymmetric case, ${\cal S}$ has been multiplied by $100$ in Fig. \ref{svsdistfig}(a) in order to
display it on the same scale as the symmetric case. As expected, ${\cal S}$ is initially small and
negative. As the amplitudes decay, the exponential terms in Eqs. (\ref{moment2}) and
(\ref{moment2a}) start to become significant when $\exp\left( {-4\alpha^2e^{-{K\over
2}(d_1+d_2)}\sin^2\phi}\right) $ becomes comparable in size to $\cos 2\phi$. This signals the
beginning of significant overlap between the two phase space regions in Fig. \ref{ESDfigb}. At this
point, there is a negative dip is ${\cal S}$, followed by an asymptotic decay back toward zero, due
to the decay of the overall $\alpha^6(d)$ dependance. The latter decay results from the regions of
Fig. \ref{ESDfigb} approaching the vacuum state at the origin. Thus, the dips occur at the point
where the entanglement starts to become unusable due to photon loss,and therefore signals the outer
limits of the distance at which the method is useful for the given input parameters.

%%%%%%%%%%%%%%%%%%%%%%%%%%
%  To a
%good approximation the large negative dips in the true entanglement witness ${\cal S}$ begin at the
%same distance at which the curves of ${\cal W}$ cross the axis; it is in this sense that the two
%eavesdropping witnesses ${\cal S}$ and ${\cal W}$ were said in the introduction to track each
%other: by plotting the graph of one function we can predict the location of features on the graph
%of the other.

%%%%%%%%%%%%%%%%%%%%%%%%%%

Note from the figure that although the large negative dip is orders of magnitudes smaller when the
decay is occurring in only one arm, it occurs at roughly twice the distance. The zero crossing of
${\cal W}$ will similarly be seen in the next section to occur at twice the distance in the
asymmetric case. This is significant because it means that the mechanism for eavesdropper detection
will work over roughly twice the distance in the asymmetric case.  As shown in Fig. \ref{ESDfigb},
the entanglement loss is slower in the asymmetric case because the two states may initially move
apart as one of them approaches the origin more rapidly than the other. In any event, as will be
seen in the next section, the dips in ${\cal S}$ occur at roughly the location where the photon
number has decayed to the point where homodyne measurements become imprecise. Thus, predictions
beyond the beginning of these dips should be considered meaningless. Henceforth, except when stated
otherwise, the figures in the remainder of this paper will be plotted for the symmetric case versus
total Alice-Bob distance, $d=d_1+d_2$; plotted this way, the asymmetric case shows only minor
differences, aside from a change of scale.

\begin{figure}%[ht]
\begin{center}
\includegraphics[width=\columnwidth]{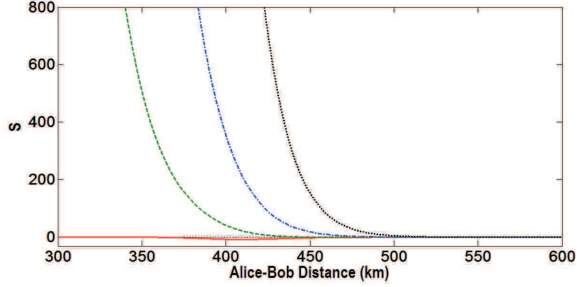}
\caption{(Color online) Behavior of entanglement witness ${\cal S}$ as a function of distance,
with and without eavesdropping, for $\alpha =1000$ and $\phi =.1$, assuming symmetric decay. The curves correspond to
no eavesdropping (solid red), $\gamma =0$ (dashed green), $\gamma =-1$ (dash-dot blue), $\gamma =-2$ (dotted black).
 }\label{Eve_svsdistfig}
 \end{center}
\end{figure}

Replacing $\hat a_2$ in Eq. (\ref{Sdef}) by the output $\hat a_{out}$ of a cloner, the effect of
eavesdropping on ${\cal S}$ can be evaluated. Examples of the results are shown in Fig.
\ref{Eve_svsdistfig}. It is clear from the plots that ${\cal S}<0$ in the absence of eavesdropping,
but switches to ${\cal S}>0$ when Eve is present.

Since ${\cal S}$ is only slightly negative at most distances, it only requires a small disturbance
to tip it to the positive side of axis. The initially large size of the positive ${\cal S}$ values
in the presence of eavesdropping may seem surprising, but it can be traced to its source: the large
value of $\langle \hat a_1^\dagger\hat a_1\rangle $ acts as a multiplier, magnifying changes in
${\cal S}$. To see this, note first that if  ${\cal S}$ is expanded out explicitly in terms of
expectation values, the only terms that change when the eavesdropper acts can be written in the
form
\begin{equation}\left( \langle \hat a_1^\dagger\hat a_1\rangle -\langle \hat a_1^\dagger \rangle^2 \right)
\langle \hat a_1^\dagger \hat a_1 \hat a_2^\dagger \hat a_2\rangle .\end{equation} The terms in the
parentheses can be written as $\langle \Delta q_1^2+\Delta p_1^2+i\left[ \hat p_1,\hat
q_1\right]\rangle$, which is nonnegative on general quantum mechanical principles; for the specific
states considered in this paper, it can be written more concretely as $\alpha^2\sin^2\phi$, which
is also clearly non-negative. Since this term is positive, ${\cal S}$ will increase if the fourth
order term multiplying it increases. With eavesdropping, the fourth-order term \emph{does} increase
by an amount proportional to $\langle \hat a_1^\dagger \hat a_1\rangle e^{-2\gamma} $, which in
turn is proportional to Alice's squared amplitude, $\alpha^2$. At small distances, ${\cal S}$ is
initially small and negative, but the amplitude $\alpha$ is large, so that this term adds a large
positive value to the entanglement witness. In more physical terms, the cloner transforms the
initial pure state en route to Bob into a mixed state, leading to a decrease in entanglement; the
effect of this loss on the witness is large because it is multiplied by the coherent state
amplitude, which we explicitly assume to be large. The loss of decoherence results from the fact
that not only are the phase space regions in Fig. \ref{ESDfigb} larger, their locations fluctuate
relative to each other about fixed average positions as a result of the uncontrolled relative phase fluctuations introduced by
the cloner.

%\begin{figure}[ht]
%\subfigure[]{
%  \includegraphics[width=\columnwidth]{long_Eve_splot}}
%\subfigure[]{
%  \includegraphics[width=\columnwidth]{Eve_splot}}
%\subfigure[]{
%  \includegraphics[width=\columnwidth]{Eve_aplot}}
%\caption{(Color online) (a) Behavior of entanglement witness ${\cal S}$ as a function of distance,
%with and without eavesdropping, for $\alpha =10^6$ and $\phi =.1$, assuming symmetric decay. The
%lower (red) line is in the absence of eavesdropping; The middle line (black) is for eavesdropping
%with $e^{-2\gamma}=2$, and the bottom line (green) has $e^{-2\gamma}={1\over 2}$.
%(b) The same as in a, but with an expanded view of the region where the curves approach the axis.
%(c) The same as in b, but for loss only on Bob's side.
% }\label{Eve_svsdistfig}
%\end{figure}

\section{An Eavesdropping Witness}\label{evewitsection}

In analogy to an entanglement witness, we wish now to introduce the concept of an
\emph{eavesdropping} witness. We will define this to be an experimentally measurable function of
the system's state which changes value in a predictable manner whenever an eavesdropper acts on the
system. Here we will introduce such a measure that will give results closely related to those of
the entanglement witness ${\cal S}$ introduced in section \ref{entwitsection}. So this new function
will also witness eavesdropping but is much easier to measure. This eavesdropping witness
${\cal W}$ is constructed from the covariance matrix of the system, and will change signs from
negative to positive at a distance that can be easily calculated. This distance changes in a
predictable manner when the system is interfered with, thus signalling the presence of an
eavesdropper.

Let $\hat q_1,  \hat p_1$ be orthogonal quadratures for beam A and
$\hat q_2,\hat p_2$ be corresponding quadratures for B. Form the vector: $\hat \eta =\left(\hat
q_1, \hat p_1 ,\hat q_2, \hat p_2\right).$ The {\it covariance matrix} $V$ is defined as the
$4\times 4$ matrix with elements $V_{ij}={1\over 2}\langle \left\{\hat \eta_i-\langle
\eta_i\rangle, \hat \eta_j-\langle \eta_j\rangle\right\}\rangle ,$ where $\left\{ ..\; ,..\;
\right\}$ denotes the anticommutator and angular brackets denote expectation value.
$V$ can be expressed in terms of three $2\times 2$ matrices as $V=\left( \begin{array}{cc}
A_1 & C\\ C^T & A_2\end{array}\right) %\label{Vmatrix}
$. $A_1$ and $A_2$ are the self-covariance matrices of each beam separately; $C$ describes
correlations between the $A_i$. An eavesdropping witness derived from the covariance matrix is then
defined as
\begin{equation}{\cal W}=1+det\; V+2 \; det \; C -det\; A_1-det\; A_2.\label{witness}\end{equation}

This function ${\cal W}$ is similar in form to an entanglement witness ${\cal W}_s$  introduced in
\cite{rsimon} and studied in detail in \cite{barbosa}, but due to the normalization differences
mentioned in the introduction, it is not the same function and so here is not a true entanglement
witness. ${\cal W}$ and ${\cal W}_s$ are in fact related by a rescaling of the quadratures, but for
the states considered in this paper ${\cal W}_s$ vanishes identically. It can be shown that for
Gaussian states, a system is entangled if and only if ${\cal W}_s<0$. ${\cal W}_s$, like ${\cal
S}$,  is based on the positive partial trace criterion \cite{peres,horod2}; however because ${\cal
W}_s$ is quadratic in the quadrature operators, it is unable to detect some forms of entanglement
that can be detected by the quartic operator ${\cal S}$. The vanishing of ${\cal W}_s$ on the
states used here is due to the fact that they are not strictly Gaussian; however we will make use
in section \ref{keysection} of the fact that the non-Gaussian terms are small for large $\alpha$.

\begin{figure}%[htb]
\begin{center}
\includegraphics[scale=.30] {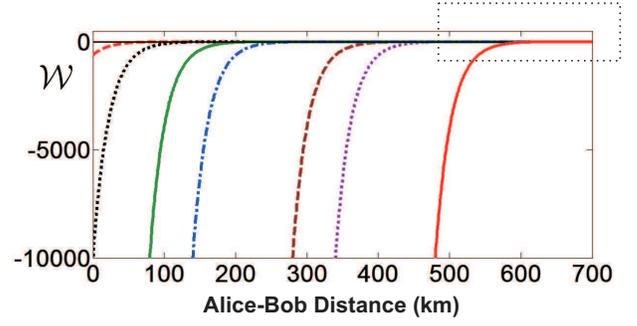}
\caption{(Color online) Eavesdropping witness ${\cal W}$ value versus Alice-Bob distance $d=d_1+d_2$.
From left to right, the curves have parameter values $|\alpha\phi|= 10$, $|\alpha\phi|= 20$, $|\alpha\phi|= 50$,
$|\alpha\phi|= 100$, $|\alpha\phi|= 500$, $|\alpha\phi|= 1000$, and $|\alpha\phi|= 5000$.
$K=.046\; km^{-1}$ is used for the $1550\; nm$ telecom window. An expanded view of the region enclosed in the dashed box is shown in Fig.
\ref{wvdistfigb} }
\label{wvdistfiga}
\end{center}
\end{figure}

Using an eavesdropping witness derived from the covariance matrix, as ${\cal W}$ is, has distinct
advantages, since $V$ is experimentally measurable via heterodyne detection and its expected
behavior with distance is straightforward to calculate. So deviations from its expected distance
dependence are easily detected. The eavesdropper's actions affect the various covariances and
moments of the states; the idea is to find a function which distills these effects into a single
number in a useful way. Clearly, many such functions are possible, but we examine here just one
example.

%: Alice and Bob simply make quadrature measurements and compare them.
%In addition, losses can be easily incorporated: annihilation operators are attenuated according to
%$\hat a_j\to t_j\; \hat a_j +\sqrt{1-t_j^2}\; \hat a_j^{(E)}$ (where $E$ denotes the vacuum or
%environment), so that if the relevant states are orthogonal then the covariance submatrices are
%affected according to \cite{barbosa}:
%$C\to C^\prime = t_1t_2C, $ and $A_j\to A_j^\prime =t_j^2\left( A_j-{1\over 4}I\right)+{1\over 4}I,$ %\label{atten}$
%where $I$ is the $2\times 2$ identity matrix. In the present case, the coherent states, being of
%finite intensity, are not orthogonal, so these expressions are approximate; the exact expressions
%are given below. (The factors of ${1\over 4}$ relative to those in \cite{barbosa} result from a
%difference in normalization used.) For propagation losses, the transmission matrices are of form
%$t_j=e^{-{1\over 2}K_j d_j}$, with  propagation distance $d_j$ in each arm. These expressions can
%be used to determine the distances over which ${\cal W}$ can become non-negative, indicating
%disentanglement.

Assuming loss rates $K_1$ and $K_2$ in each arm,  then the covariance matrix is
\begin{eqnarray}\label{matrixdiss}V&=&\left(
                                        \begin{array}{cc}
                                          A_1^\prime & C^\prime \\
                                          C^{\prime\; T} & A_2^\prime \\
                                        \end{array}
                                      \right)
                                      =\left(
                                         \begin{array}{cccc}
                                           a_1^\prime  & 0 & b^\prime & 0 \\
                                           0 & a_1^\prime & 0 & c^\prime \\
                                           b^\prime & 0 & a_2^\prime & 0 \\
                                           0 & c^\prime & 0 & a_2^\prime \\
                                         \end{array}
                                       \right) ,\end{eqnarray}
                                       where
\begin{eqnarray}& & a_j^\prime (d_j)= {{|\alpha |^2}\over 2}\left( |N|^2f(\theta ,\phi ,d_j)
        -1\right)e^{-K_jd_j} +{1\over 4}\label{aeq} \\
& & b^\prime (d_1,d_2) = {{|\alpha |^2}\over 2}\left( |N|^2g(\theta ,\phi ,d_1,d_2) \right. \label{beq}\\
& & \qquad \qquad \qquad\qquad  -\left. \cos 2\phi \right)e^{-{1\over 2}(K_1d_1+K_2d_2)}\nonumber\\
& & c^\prime (d_1,d_2) = {{|\alpha |^2}\over 2}\left( |N|^2g(\theta ,\phi ,d_1,d_2)\right. \label{ceq}\\
& & \qquad \qquad \qquad\qquad\qquad  \left.  - 1 \right)  e^{-{1\over 2}(K_1d_1+K_2d_2)}\nonumber\end{eqnarray}
with $j=1,2$. Here we have also defined \begin{eqnarray}f(\theta ,\phi ,d_j)&=&\left[ 1+\cos 2\phi
\cos \theta e^{-4|\alpha |^2\sin^2\phi\; e^{-Kd_j}}\right]\\ g(\theta ,\phi ,d_1,d_2)&=&\left[ \cos
2\phi \right. \\
& & \quad \left. +\cos \theta e^{-4|\alpha |^2\sin^2\phi \; e^{-(K_1d_1+K_2d_2)/2}}
\right]\nonumber .\end{eqnarray} The values of $a^\prime,\; b^\prime,\; c^\prime $ at zero distance
will be denoted by $a,\; b,\; c$. Distance dependance also arises through $N(d_1,d_2)$. The
entanglement witness is then
\begin{eqnarray}& & {\cal W} = 1+(a_1^\prime a_2^\prime )^2 +( b^\prime c^\prime )^2-(b^{\prime\; 2}+c^{\prime\; 2})a_1^\prime a_2^\prime \nonumber \\
& & \qquad \qquad +2b^\prime c^\prime- a_1^{\prime\; 2}-a_2^{\prime \; 2}.
\end{eqnarray}
%Eq. \ref{matrixdiss} is in the standard normal form for a two-mode Gaussian state; the entries
%$a_1^\prime$, $a_2^\prime$, $b^\prime$, and $c^\prime$ are all simply related to symplectic
%invariants \cite{giedke,olivares}. The resulting Gaussian Wigner function is given by
%\begin{equation}W(\bm \eta) ={1\over {\pi^2\sqrt{det(V)}}} e^{-{1\over 2}\left( \bm\eta -\langle \bm\eta \rangle\right)^T V^{-1}
%\left( \bm\eta -\langle \bm\eta \rangle\right)} .\end{equation}

\begin{figure}%[htb]
\begin{center}
\includegraphics[scale=.30] {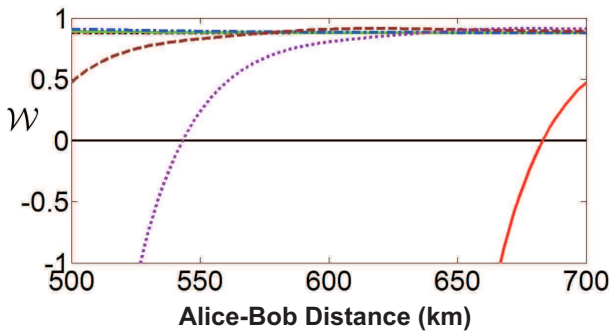}
\caption{(Color online) Expanded view of the region enclosed in the dashed box in Fig. \ref{wvdistfiga}.
The curves have parameter values $|\alpha\phi|= 500$(dashed brown), $|\alpha\phi|= 1000$ (dotted violet), and $|\alpha\phi|= 5000$ (solid red).
$K=.046\; km^{-1}$ is used for the $1550\; nm$ telecom window. }
\label{wvdistfigb}
\end{center}
\end{figure}

Henceforth we assume that in both channels the rate for fiber loss is that of the $1550\; nm$
telecom window, $K_1=K_2\equiv K=.046\; km^{-1}$, corresponding to $3$ dB loss per $15$ km. We also
write now for the most part express results in terms of the total Alice-to-Bob distance, $d=d_1+d_2$. In this manner, the symmetric case (equal travel distances in both channels, $d_1=d_2$) and the case where Alice generates the state in her lab ($d_1=0$, with no losses on her side) can both be
expressed in a unified manner.
Plots of ${\cal W}$ vs. distance are given in Fig. \ref{wvdistfiga}. ${\cal W}$
starts with large negative values at $d=0$ and its magnitude decays rapidly with distance due to
propagation losses. Close inspection shows that ${\cal W}$ crosses from negative to positive values
at finite distances (see the expanded version in Fig. \ref{wvdistfigb}).

The exponential terms in $f$ and $g$ are negligible except at large distances, by which point the
$\alpha^2(d)$ terms that multiply them in Eqs. (\ref{aeq})-(\ref{ceq}) have decayed away. As a result,
these terms can be neglected for most purposes. Dropping them, it is then seen that all of the
curves in Fig. \ref{wvdistfigb} converge to a common asymptote as $|\alpha |\to \infty$, located at
${\cal W}=(1-a^2)^2= \left( {{15}\over {16}}\right)^2\approx 0.8789.$

\section{Crossing Thresholds}\label{thresholdsection}

Entanglement sudden death (ESD) is the sudden loss of entanglement in finite time---corresponding
here to finite distance---in contrast to the more common asymptotic loss of entanglement due to
decoherence \cite{yu,almeida,ann}. Although, as mentioned, the eavesdropping witness ${\cal W}$ is
not an entanglement witness, behavior analogous to ESD occurs here. The point at with the axis is
crossed moves in the presence of eavesdropping and closely tracks features of the true entanglement ${\cal S}$
witness discussed in Sec. \ref{entwitsection}; as a result, the location of this crossing point can be used
as means of eavesdropper detection.

\begin{figure}%[htb]
\begin{center}
\includegraphics[scale=.30]{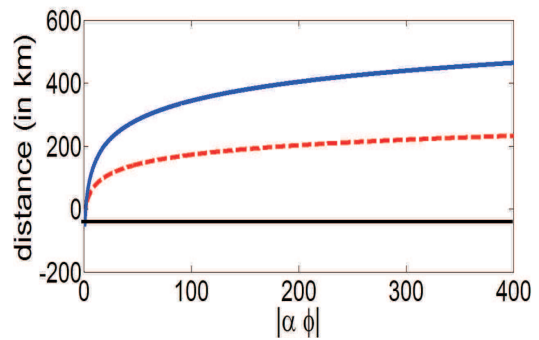}
\caption{(Color online) The distance $d_0$ at which axis crossing occurs, as a function of the parameter
$ |\alpha\phi |\approx |\alpha\sin\phi |$ for $K=.046\;km^{-1}$. The solid blue curve is the Alice-to-Bob distance.
This the same as the source-to-Bob distance for the case of loss in only one arm (source in Alice's lab), and is double the source-to-Bob distance for case of equal distances in both branches (dashed red curve). }
\label{ESDfiga}
\end{center}
\end{figure}

For $\phi =0$ the matrix
elements reduce to $a={1\over 4}$ and $b=c=0$, so we find that ${\cal W}
=1-a^4-2a^2=(1-a^2)^2=\left({{15}\over {16}}\right)^2>0,$ at all distances. But for nonzero $\phi$,
${\cal W}$ changes sign when $|\alpha (d)| =\sqrt{{15}\over 4}\; \csc\phi $. Solving for distance,
we find that the sign change occurs when the distance between Alice and Bob is
\begin{equation}d_0={2\over K}\ln \left( \sqrt{8\over {15}}\alpha
\sin\phi\right) .\label{dnoeve}\end{equation} These results are plotted
in Fig. \ref{ESDfiga}. Although here we restrict ourselves to small $\phi$, it may be noted in
passing that, for fixed $\alpha$, the crossing distance is largest at $\phi={\pi\over 2}$, i.e.
when the entangled states are $|\alpha\rangle$ and $|-\alpha\rangle$. As the distance formula makes
clear, crossing can always be made to occur at any distance desired by choosing appropriate values
of $\phi$ and $\alpha$.

\begin{figure}%[htb]
\begin{center}
\includegraphics[scale=.30]{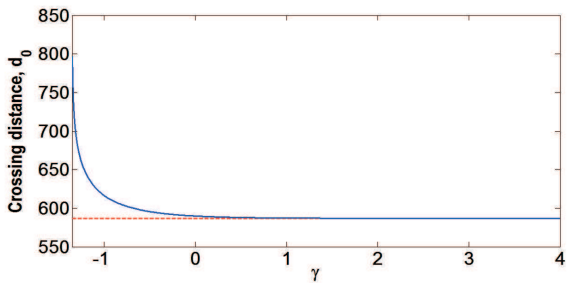}
\caption{(Color online) The solid blue line is the distance $d(\gamma)$ (in kilometers) between Alice and Bob at which
${\cal W}$ crosses the axis, as a function of eavesdropping parameter
$\gamma $. The amplitude and phase values assumed are $\alpha=10^5$ and
$\phi =.1$. The dashed red line shows the distance $d_0$ in
the absence of eavesdropping.}
\label{crossvsgammafig}
\end{center}
\end{figure}

\begin{figure}%[htb]
\begin{center}
\includegraphics[width=\columnwidth]{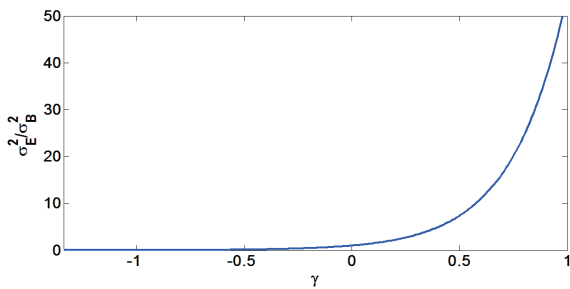}
\caption{(Color online) The ratio of added variances
for Bob and Eve, $r={{\sigma_E^2}\over {\sigma_B^2}}$ is plotted versus $\gamma$, for $\phi=.1$. The curve is
independent of $\alpha$.}
\label{crossratiofig1}
\end{center}
\end{figure}

Let us now consider the effect of eavesdropping on ${\cal W}$.
%Alice's measurements are unchanged
%and from the expressions of Section \ref{clonesection} it is readily verified that the
%cross-correlations between Alice and Bob are unchanged. So the submatrices $A_1$ and $C$ are
%unaltered.  But
The variances on the diagonal of $A_2$ are increased by ${1\over 4}e^{-2\gamma}$, so the crossing
distance is now altered in the presence of eavesdropping to the new value
\begin{equation}d(\gamma )={2\over K}\ln \left[ \sqrt{8\over {15}}\left( 1-{1\over{15}}e^{-2\gamma}
\right)^{-1}\alpha^2 \sin^2\phi \right] .\label{dwitheve}\end{equation} $d(\gamma )$ becomes
complex for $\gamma<\gamma_0$, where $\gamma_0\equiv -{1\over 2}\ln 15\approx -1.3540$. So for
eavesdropping parameters below $\gamma_0$ there is no crossing, and ${\cal W}$ is always negative.
This lack of axis-crossing provides a clear and unambiguous signal of eavesdropping. For
$\gamma>\gamma_0$, the crossing distance becomes finite, starting at large values and decaying
rapidly to $d_0$ as $\gamma$ increases (Fig. \ref{crossvsgammafig}). Since the ratio of Eve's added
variance (beyond the vacuum value) to Bob's added variance, $r={{\sigma_E^2}\over {\sigma_B^2}}
=e^{4\gamma}$, increases exponentially with $\gamma$, the shift in crossing point is large (or
infinite) for parameter values where Eve can measure the quadratures with precision (large negative
$\gamma$). The shift only becomes too small to detect exactly in the region where Eve's variance is
too large for her to extract an accurate measurement (positive $\gamma$). This is illustrated in
Fig. \ref{crossratiofig1}, where $r$ is plotted versus $\gamma$. By the time the shift in crossing
point is reduced to $1$ meter in size, Eve's variance is $2.2$ times that of Bob; by the time $\Delta d$ drops to $.5$
meters, the added variance ratio grows to $r=8.6$.

\begin{figure}%[htb]
\begin{center}
\includegraphics[width=\columnwidth]{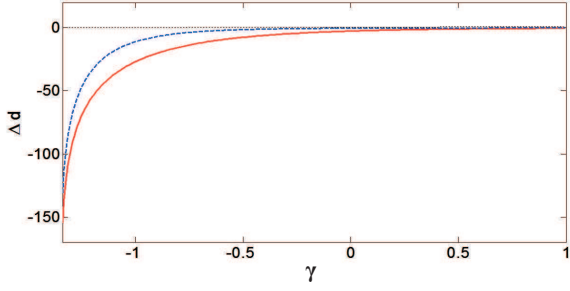}
\caption{(Color online) The change $\Delta d$ in the distance at which the curve of ${\cal W}$ crosses the value ${\cal W}$ is plotted versus the eavesdropping parameter, $\gamma$, for the values $\Lambda =-1$ (dashed blue) and $\Lambda =-10$ (solid red). The curves are independent of $\alpha$ and change very little for $\Lambda<-10$. The value $\phi=.1$ was used for the plots. }
\label{crossfig1}
\end{center}
\end{figure}

The average number of photons in a coherent state is related to the amplitude by $\langle n\rangle
=\alpha^2$, so if the amplitude is decaying as $\alpha (d)=\alpha e^{-Kd/2}$, then the distance
$D_1$ at which the number of photons decays to roughly one is
\begin{equation}D_1={2\over K}\ln\alpha .\end{equation} More generally, the distance at which the
number has decayed to $\langle n\rangle =N$ is \begin{equation}D_N={2\over K}\ln\left( {\alpha\over
\sqrt{N}}\right) .\end{equation} Unless $\phi $ is relatively large (of order $.1$ or more), the
points at which the curves cross the axis tend to be in the regions where a small number of photons
remain in the beam, making homodyne or heterodyne measurements at those points imprecise. As a
result, it is advantageous instead to look at the distances at which the ${\cal W}$ curve crosses
some negative value $\Lambda$, instead of the distance where it crosses zero. Let the distance at
which ${\cal W}=\Lambda$ be $d(\gamma,\Lambda)$. In the absence of an eavesdropper, the distance
would be $d(\infty ,\Lambda )$, so that the distance that this crossing moves in the presence of
eavesdropping is $\Delta d(\gamma,\Lambda) =d(\gamma,\Lambda)-d(\infty ,\Lambda ).$ It is
straightforward to show that
\begin{eqnarray}d(\gamma,\Lambda) &=& {2\over K}\ln \left[ {{\alpha^2\sin^2\phi }\over
{ F(\gamma )-{\Lambda \over {F(\gamma )}}}}\right]\\
\Delta d(\gamma,\Lambda) &=& {2\over K} \ln \left[ {{{{15}\over {16}}-{{16}\over {15}}\Lambda}\over
{ F(\gamma )-{\Lambda \over {F(\gamma )}}}} \right] ,
\end{eqnarray} where $F(\gamma )\equiv {{15}\over {16}} \left( 1-{1\over {15}} e^{-2\gamma}\right) $.
This shift is independent of the initial value of $\alpha$, and varies only very slowly with $\Lambda$.
The value of $\Lambda $ used can be
chosen as appropriate for the given experiment to ensure that there are still sufficient numbers of
photons remaining in the beam for accurate homodyne measurements. The size of this shift for the
particular values $\Lambda =-1$ and $\Lambda =-10$ is shown in Fig. \ref{crossfig1}. For more negative values of $\Lambda $, the curves are nearly indistinguishable from that of $\Lambda =-10$.

\begin{figure}%[htb]
\begin{center}
\includegraphics[scale=.30]{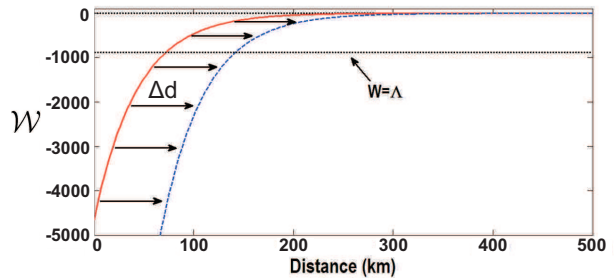}
\caption{(Color online) The curves shift horizontally by approximately a constant amount
$\Delta d(\gamma,\Lambda)$ in the presence of eavesdropping. Here the solid red line is in the absence of
eavesdropping for $\alpha=1000$ and $\phi=.1$. The dashed blue line is in the presence of eavesdropping
with $\gamma =-1$.
The crossing of the ${\cal W}=\Lambda$ line can be used instead of the ${\cal W}=0$ crossing; this allows more
photons to still be present for measurement, increasing measurement accuracy.
} \label{shiftingfig}
\end{center}
\end{figure}

%\begin{figure}%[htb]
%\begin{center}
%\includegraphics[width=\columnwidth]{crossingvsvar1}
%\caption{(Color online) Top: The change $\Delta d$ in distance at which the curve of ${\cal W}$ crosses the axis
%is plotted versus the eavesdropping parameter, $\gamma$. Bottom: the ratio of added variances
%for Bob and Eve, $r={{\sigma_E^2}\over {\sigma_B^2}}$ is plotted versus $\gamma$. Both plots are
%given for $\alpha =10^5$ and $\phi=.1$.}
%\label{crossratiofig1}
%\end{center}
%\end{figure}

\section{Information and Secret Key Rate}\label{keysection}

Although the primary goal in this paper is to use entanglement in the phase in order to detect
eavesdropping on a classically modulated channel, rather than to use the entangled phase for
encryption or encoding itself, we briefly consider here other possibilities which are available in case a full quantum key distribution is desired.

In particular, the same setup can be used to generate a key from the homodyne measurements themselves. The
possible phase values measured by each participant can be divided up into bins and the bin in which
a measurement falls then determines a value for the key. In this situation, the mutual information
between the participants and the eavesdropper is relevant to determining if it is possible to
distill a secret key. With a sufficient number of bins, the phase variable can still be treated as
approximately continuous.

%\begin{figure}%[htb]
%\begin{center}
%\includegraphics[scale=.30]{crossingvsvar2}
%\caption{(Color online) The shift in crossing distance $\Delta d$ and the added variance ratio $r$ from the
%previous figure are plotted against each other. It can be seen that when Eve is capable of measuring
%the quadratures with precision (small $r$), the shift is large and easily detectable (large $\Delta d$).}
%\label{crossratiofig2}
%\end{center}
%\end{figure}

The secret key rate is given by
\begin{equation}\kappa =I(A:B)-I(B:E),\label{keydef}\end{equation} where $I(A:B)$ and $I(B:E)$ are respectively the
mutual information between Alice and Bob and between Bob and Eve. The mutual information is simply
the difference between the von Neumann entropies of the individual subsystems and the total
two-beam system, $S_{vn}=-\mbox{Tr}\left[\rho\ln\rho\right]$; for example,
$I(A:B)=S_{vn}(\rho_A)+S_{vn}(\rho_B)-S_{vn}(\rho_{AB}).$ If $K>0$, then it is possible to distill
a secret key via privacy amplification. If the difference in Eq. (\ref{keydef}) is negative, then $\kappa$ is taken to be zero.
The mutual information can be calculated numerically from
the density operator of the system. However an approximate but simpler and more transparent
evaluation can be obtained by noting that the system in question can be treated as an approximately
Gaussian system for small $\phi$. This can be seen, for example by calculating the characteristic
function (the Fourier transform of the Weyl operator) or the Wigner function of the system. The
characteristic function for example, is of the form
\begin{eqnarray}& & \gamma (\lambda ,\zeta )= {1\over 2} e^{-{1\over 4}\left( |q_1+i\zeta_1|^2+|q_2+i\zeta_2|^2\right)
}\\ & & \;\;\times
\int d^2\lambda\; d^2\chi \; e^{-\left( 2|\alpha |^2+|\lambda |^2 +|\chi |^2\right)}\nonumber \\
&& \;\;\times  e^{{i\over 2}\left[ \left( q_1+i\zeta_1\right)\lambda^\ast +\left( q_2+i\zeta_2\right)\chi^\ast
+\left( q_1-i\zeta_1\right)\lambda +\left( q_2-i\zeta_2\right)\chi\right]}\nonumber \\
& & \; \;\times\left( e^{\alpha \left( \lambda_r+\chi_r\right) \cos\phi +2\alpha (\lambda_i-\chi_i)\sin\phi} \right.\nonumber \\
& & \qquad\quad +
e^{\alpha \left( \lambda_r+\chi_r\right) \cos\phi +2\alpha (\chi_i-\lambda_i)\sin\phi }\nonumber \\
& & \qquad\quad +e^{\alpha \left( \lambda_r+\chi_r  \right)\cos\phi +2i\alpha
(\chi_r-\lambda_r)\sin\phi}\nonumber \\ & & \qquad\quad +\left. e^{\alpha \left(
\lambda_r+\chi_r\right) \cos\phi +2\alpha (\chi_r-\lambda_r)\sin\phi}\right)\nonumber
\end{eqnarray} Here, subscripts $1$ and $2$ label Alice's and Bob's sides, while subscripts
$r$ and $i$ label real and imaginary parts. Because of the terms in the last large parentheses,
$\gamma$ is a sum of four Gaussians. But when $\phi$ is small, the sine terms in the exponentials
become negligible compared to the cosine terms, leaving all four of these terms equal. The only
case when this argument breaks down is when the differences $\chi_i-\lambda_i$ or
$\chi_r-\lambda_r$ are large; however this part of the integration range is strongly suppressed by
the term $e^{-\left( 2|\alpha |^2+|\lambda |^2 +|\chi |^2\right)}$ in the second line. Thus, to a
high degree of accuracy, we can treat the system as Gaussian. This approximation becomes better as
the distance becomes large and the amplitudes decay to small values, which is exactly the region of
greatest interest to us. We therefore compute all information-related quantities in the Gaussian
approximation.

\begin{figure}%[htb]
\begin{center}
\includegraphics[scale=.30]{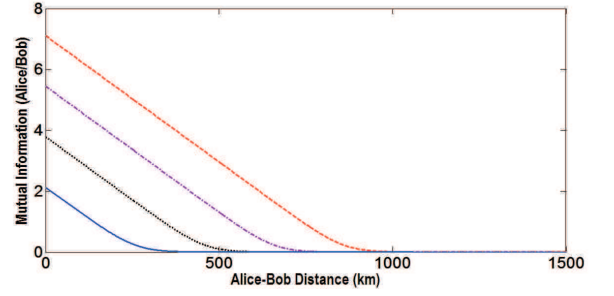}
\caption{(Color online) Mutual information between Alice and Bob, assuming both
have same initial amplitude $\alpha$.
From the top line downward, the initial amplitudes are $\alpha=10^5$ (red), $\alpha=10^4$
(violet), $\alpha =10^3$ (black), and $\alpha =100$ (blue). $\phi=.1$ for all curves.}
\label{symmfig}
\end{center}
\end{figure}

For a two-mode Gaussian state, the mutual information can be obtained directly from the covariance
matrix. Define the binary entropy function $h(x)=(x+{1\over 4})\ln (x+{1\over 4})+(x-{1\over 4})\ln
(x-{1\over 4})$ and the discriminant of the covariance matrix $\Delta =det(A)+det(B)+2det(C)$. Then
the quantum mutual information is \cite{serafina,olivares}:
\begin{equation}I(A:B)=h(\sqrt{Det(A)})+h(\sqrt{Det(B)}) -h(d_+)-h(d_-),\end{equation}
where the symplectic eigenvalues of the covariance matrix are
\begin{equation}d_\pm=\sqrt{{\Delta \pm \sqrt{\Delta^2-\sqrt{\Delta^2-4\; Det(V)}}}\over 2}.\end{equation}

\begin{figure}%[htb]
\begin{center}
\includegraphics[scale=.30]{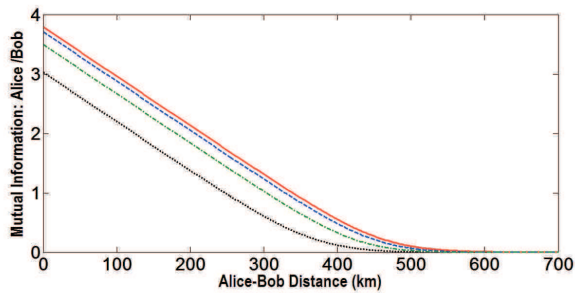}
\caption{(Color online) Mutual information between Alice and Bob in the presence of eavesdropping,
assuming they have equal initial amplitudes and equal losses.  Solid red curve: no eavesdropper.
Dotted black: $\gamma =-1$. Dash-dot green: $\gamma =0$. Dashed blue: $\gamma =1$. The values $\phi=.1$
and $\alpha =1000$ were used for all curves. The same curves give the mutual information between Bob and Eve, but with $\gamma$ and $-\gamma$ interchanged.}
\label{eveinfosymmfig}
\end{center}
\end{figure}

\begin{figure}%[htb]
\begin{center}
\includegraphics[scale=.30]{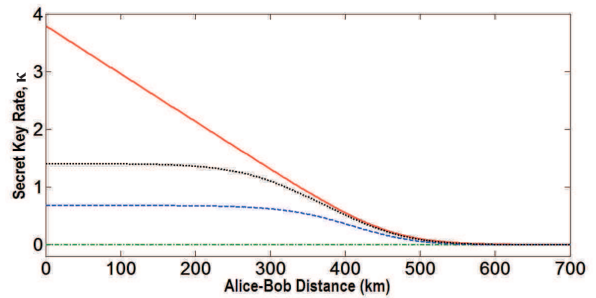}
\caption{(Color online) Secret key rate $\kappa$ between Alice and Bob in the presence of eavesdropping,
assuming they have equal initial amplitudes and equal losses.  Solid red curve: no eavesdropper.
Dotted black: $\gamma =2$.  Dashed blue: $\gamma =1$. Dash-dot green: $\gamma =0$. $\kappa
$ vanishes identically for all
$\gamma\le 0$. The values $\phi=.1$
and $\alpha =1000$ were used for all curves.}
\label{keysymmfig}
\end{center}
\end{figure}

Plots of the mutual information between Alice and Bob in the absence of eavesdropping in Fig.
\ref{symmfig}. In the presence of eavesdropping, examples are shown in Fig. \ref{eveinfosymmfig}.
As would be expected, the mutual information they share decreases as $\gamma$ decreases, i.e. as
Bob's variance increases and Eve's drops. Because of the relation between Bob's variance and Eve's,
it can be noted that the mutual information between Alice and Eve is given by the same formula, but
with the sign of $\gamma$ reversed. This makes calculating the secret key rate very simple, and
leads to results such as those shown in Fig. \ref{keysymmfig}. The key rate remains positive as
long as $\gamma>0$, which is is equivalent to saying $\sigma_E^2>\sigma_B^2$. It should be noted
that the distances at which the information approaches zero are roughly equal to the distances at
which ${\cal S}$ became small in Sec. \ref{entwitsection}. Since $\gamma =0$ corresponds to
$\sigma_E^2 ={1\over 4}$, it follows that the maximum allowed noise in the system for arrangement to remain secure
is $\sigma^2_{noise}<{1\over 4}$.

%The maximum allowable noise variance allowed while maintaining security is clearly
%$\sigma_{noise}^2={1\over 4}e^{-2\gamma}.$

\begin{figure}%[htb]
\begin{center}
\includegraphics[scale=.30]{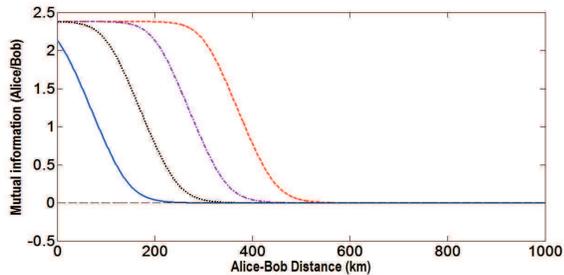}
\caption{(Color online) Mutual information between Alice and Bob, assuming they have different initial
amplitudes $\alpha$ and $\beta$.
From the right to left, Bob's initial amplitudes are $\beta=10^5$ (red), $\beta=10^4$ (violet), $\beta =10^3$
(black), and $\beta =100$ (blue). $\phi=.1$  and $\alpha =100$ for all curves.}
\label{asymmfig}
\end{center}
\end{figure}

As an interesting aside, up to this point, although different amounts of loss were allowed in
Alice's and Bob's channels due to different propagation distances, it has always been assumed that the initial amplitudes were equal for both
lines. If we allow different initial amplitudes $\alpha$ and $\beta$, respectively, for Alice and
Bob, then the information decreases more slowly with distance (Fig. \ref{asymmfig}). The reason for
this is similar to the explanation given earlier (see Fig. \ref{ESDfigb}) for the
greater distance in the presence of asymmetric decay.

\section{Conclusions} \label{conclusion}
We have analyzed effects of loss and eavesdropping in a system for distributing key bits via entangled
coherent states over long distances. We have demonstrated that when combined with the entanglement-witness or eavesdropping-witness approach,
the entangled coherent state scheme described here can in principle be used to detect eavesdropping over distances on the order of hundreds of kilometers.

Besides differing conceptually from previous approaches, our results for coherent-state QKD based on the
use of an entanglement and eavesdropping witnesses for eavesdropper detection offers distinct advantages over
use of a Bell-type inequality for that purpose. In particular, comparing the above results with those in
\cite{kirby}, we see that sign changes of ${\cal W}$ always occur at larger distances than the loss of Bell non-locality
resulting from the same  external interventions on the coherent states induced. Hence,
${\cal W}$, as well as ${\cal S}$, is available for eavesdropping detection over larger distances than is the Bell-type inequality of
the proposal on \cite{kirby}, extending the range of distances in which the phase-entangled coherent states are
known to be useful for QKD: simulations in \cite{kirby} showed the Bell inequality method to be useful up to distances on the order of tens of kilometers, while the method discussed below has promise to extend the range to the order of several hundred kilometers.
Moreover, the entanglement witness method requires only a single trigger photon, rather
than the triple-coincidence trigger required for testing the Bell-type inequality, a substantial
practical improvement.

Of the two eavesdropping witnesses, one (${\cal W}$) is straightforward to implement experimentally, while the
other (${\cal S}$) provides a rigorous measure of entanglement loss in the presence of eavesdropping. The
question remains as to whether there is some other measure that provides both features for this
system: a true entanglement witness that is readily accessible experimentally. It would be of
particular interest to find a {\it strong} entanglement witness that would serve this purpose. In
any case, the general idea of using an entanglement witness or some related function as an
eavesdropping witness or quantum tripwire for eavesdroppers can certainly be exported to
communication systems beyond the specific entangled coherent state system considered here.

Finally, we have shown that the method is potentially useful up to distances of hundreds of kilometers, in
contrast to methods based on single-photon communication which are restricted to distances of tens of kilometers at
most. It remains to be seen if the method may be combined with the use of quantum repeaters \cite{briegel} in order to
extend the working distance to even greater lengths.

\section*{Acknowledgements}
This research was supported by the DARPA QUINESS program through US Army Research Office award
W31P4Q-12-1-0015. We would like to thank Prof. N. L\"utkenhaus for a very helpful discussion.

\end{document}